# Electronic structure reorganization in MPS$_3$ via d-shell-selective alkali metal doping

Jonah Elias Nitschke[1*], Preeti Bhumla[2], Till Willershausen[1], Patrick Merisescu[3], David Janas[1], Lasse Sternemann[1], Michael Gutnikov[1], Karl Schiller[1], Valentin Mischke[1], Michele Capra[1], Mira Sophie Arndt[1], Silvana Botti[2], Mirko Cinchetti[1**]

[1] TU Dortmund University, Otto-Hahn-Straße 4, 44227 Dortmund, Germany

[2] Research Center Future Energy Materials and Systems of the University Alliance Ruhr and Interdisciplinary Centre for Advanced Materials Simulation, Ruhr University Bochum, Universitätsstraße 150, D-44801 Bochum, Germany

[3] University of Bath, Clavertown Down, Bath BA2 7AY, United Kingdom

*E-mail: jonah.nitschke@tu-dortmund.de

**E-mail: mirko.cinchetti@tu-dortmund.de

## Abstract

Semiconducting two-dimensional (2D) antiferromagnetic (AFM) transition-metal thiophosphates (MPS$_3$) offer promising opportunities for spintronic applications due to their highly tunable electronic properties. While alloying and intercalation have been shown to modulate ground states, the role of d-shell filling in governing these transitions remains insufficiently understood. Here, we investigate electron doping effects in MPS$_3$ using angle-resolved photoemission spectroscopy (ARPES), X-ray photoelectron spectroscopy (XPS), and density functional theory (DFT+U). Lithium and cesium deposition are employed to induce doping across different MPS$_3$ compounds. We identify two distinct doping mechanisms: in MnPS$_3$, electrons are primarily donated to the P$_2$S$_6$ ligand clusters, with negligible Mn 2p core-level shifts and no major changes in the valence band. In contrast, FePS$_3$, CoPS$_3$, and NiPS$_3$ exhibit clear reductions in transition-metal oxidation states, with a ~1.0 eV reduction in spin-orbit splitting for Co upon doping. ARPES on CoPS$_3$ reveals a ~400 meV shift of Co-derived bands towards higher binding energies and new dispersive states up to 1 eV above the valence band maximum, indicating metallic behavior. These results establish a direct correlation between d-shell filling and doping response, highlighting alkali metal doping as a tunable route to tailor the electronic and magnetic properties of 2D AFM semiconductors for spintronic applications.

# Introduction

Van der Waals (vdW) materials with intrinsic magnetic order have gained significant attention due to their potential for novel quantum technology applications[1,2]. Their two-dimensional (2D) nature and ease of exfoliation[3,4] make them extremely versatile, enabling a variety of tuning strategies, including external gating[5,6], alloying[7], intercalation[8,9], and engineered stacking of different layers to create designed interfaces that exploit moiré potentials and proximity effects[2].

Among 2D vdW magnets, antiferromagnetic (AFM) compounds are gaining much attention due to their intrinsic insensitivity to external magnetic fields and ultra-fast spin dynamics in the terahertz regime[10]. A particularly interesting family of 2D AFMs are the transition-metal thiophosphates ($MPS_3$), whose magnetic structure arises from a honeycomb lattice of transition metal ions ($M^{2+}$ = $Fe^{2+}$, $Ni^{2+}$, $Co^{2+}$, $Mn^{2+}$) within the crystal field of surrounding $[P_2S_6]^{4-}$ bipyramidal units[11–15] (**Figures 1a and 1b**). In these systems, magnetism originates from competing exchange interactions, primarily between transition-metal cations via direct and superexchange pathways[16]. These interactions create a delicate balance, which makes $MPS_3$ compounds highly tunable through internal modifications, such as alloying and intercalation, as well as external stimuli like strain and electrostatic gating[17,18].

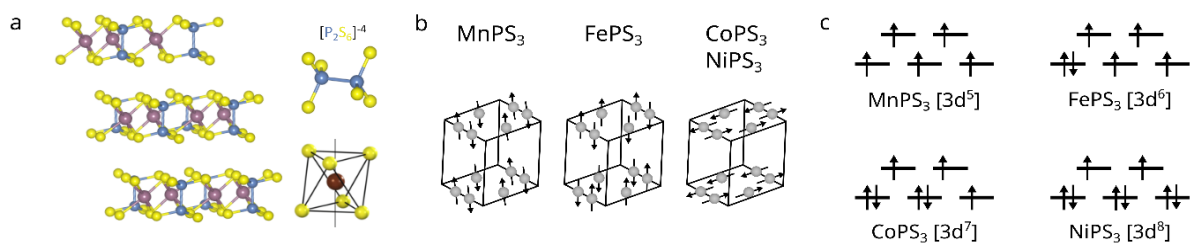

**Figure 1. Structural, magnetic, and electronic properties of $MPS_3$ compounds.** (a) Crystal structure of $MPS_3$ showing the shifted stacking parallel to the cleavage plane on the left side. The right side displays the structure of the $P_2S_6$ clusters along with the octahedral environment around the M ions by the surrounding S atoms. (b) Varying antiferromagnetic spin structures and (c) d-orbital electronic configuration of the d-orbitals of the different $MPS_3$ materials.

Recent efforts have demonstrated that alloying in $MPS_3$, through the substitution of transition-metal cations, can significantly alter exchange interactions[7,19,20]. This leads to changes in the magnetic ground state, including transitions between Néel, zigzag, ferrimagnetic and spin-glass orders[21–23]. Intercalation with molecular species or electrochemically inserted cations, on the other hand, can modify the balance between direct and superexchange interactions, inducing phase transitions from antiferromagnetic to ferromagnetic or ferrimagnetic states[8].

A common consequence of both alloying and intercalation is the alteration of d-shell filling, either by introducing charge carriers or by substituting the transition-metal species. Such modifications have a direct impact on magnetic exchange pathways and ordering, yet a comprehensive understanding of how d-electron count governs the emergence of magnetic and electronic phases remains lacking. This gap in understanding limits the controlled design of $MPS_3$-based materials for future quantum technologies.

Here, we combine angle-resolved photoemission spectroscopy (ARPES), x-ray photoelectron spectroscopy (XPS), and DFT+U calculations to investigate how electron doping via lithium and cesium deposition affects the electronic structure across the MPS$_3$ family. Our systematic study reveals compound-specific responses, with MnPS$_3$ exhibiting markedly different behavior compared to FePS$_3$, CoPS$_3$, and NiPS$_3$. We aim to understand the underlying mechanisms governing these differences and their implications for electronic and magnetic structure modification. In contrast to previous experimental and theoretical works on alkali metal intercalation[21,24–28], we focus specifically on how electron doping affects the d-shell filling of the M$^{2+}$ ions (**Figure 1c**), and how alteration of the d-shell filling influences the electronic structure of the compounds.

In the following we highlight MnPS$_3$ and CoPS$_3$ as two representative cases that exemplify the distinct doping responses observed, while detailed results for FePS$_3$ and NiPS$_3$ can be found in the Supplementary Information. By employing XPS, we investigate the changes induced to the MPS$_3$ core-level state due to Lithium doping. This reveals two different doping mechanisms that either lead to a change of oxidation state of the M$^{2+}$ ion (for M = Fe, Ni, Co) or a predominant electron localization on the P$_2$S$_6$ clusters (for M = Mn). Complementary ARPES measurements on the bare MPS$_3$ surfaces are first used to benchmark our DFT+U calculations, before the investigation of the Li doped surface uncovers how the electron doping reshapes the valence band structure, most notably in CoPS$_3$ with the least stable d-shell configuration. These findings are corroborated by further cesium based measurements, revealing that the induced modifications scale with the doping level and are therefore highly tunable.

Together, these findings establish alkali metal doping as a powerful and selective strategy to modulate the electronic structure of 2D antiferromagnets. By directly linking d-shell occupancy to band structure evolution and charge redistribution, our work offers a clear framework for tailoring the electronic and magnetic functionalities of MPS$_3$ compounds, paving the way for their integration into next-generation spintronic devices.

# Results

## XPS study

To assess how the electronic configuration of the transition metal ions evolves with electron doping, we investigate the M 2p core-level spectra shown in **Figures 2a–d.** These spectra allow us to identify changes in oxidation state and multiplet structure arising from lithium doping. The data reveal two distinct response categories: one exemplified by $MnPS_3$ and the other by Fe-, Co-, and $NiPS_3$.

Starting with $MnPS_3$ (**Figure 2a**), the undoped surface shows sharp spin-orbit split $2p_{3/2}$ and $2p_{1/2}$ peaks, both exhibiting pronounced multiplet splitting due to interactions between core and valence electrons[29–35]. This structure is well reproduced using multiplet-based fitting models from Gupta et al. [29,30] (see **Figure S1**). The 2p peaks are followed by well-defined satellite features attributed to shake-up processes[34,36]. After lithium doping, the Mn 2p spectrum remains largely unchanged: no significant shift is observed in the main peaks, and only slight peak broadening and a reduction in satellite intensity occurs. This suggests that Mn retains its 2+ oxidation state.

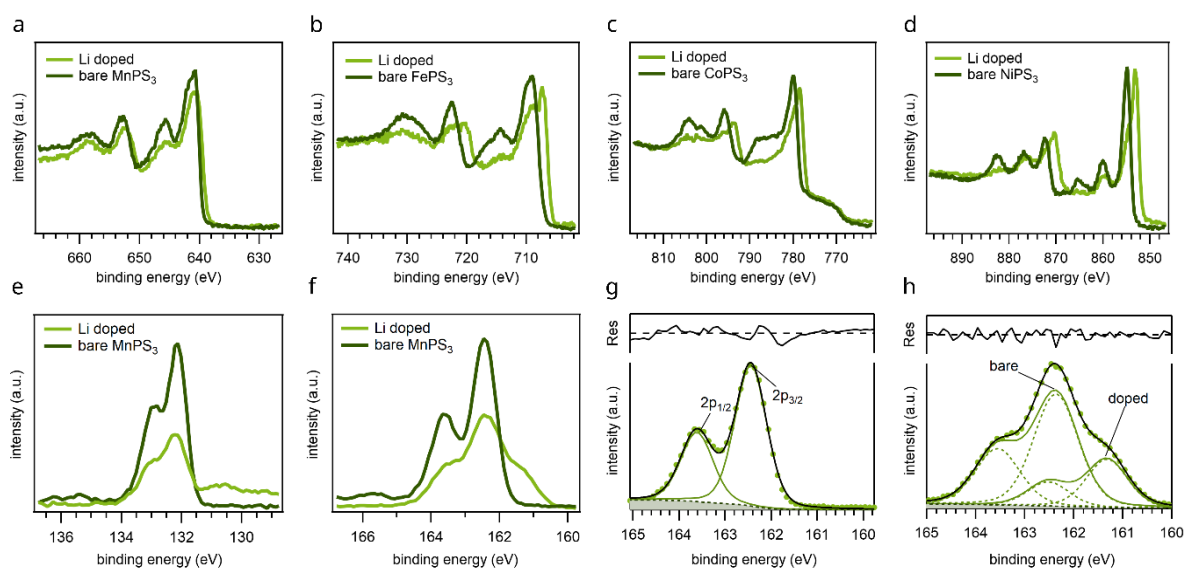

**Figure 2. XPS spectra for M 2p, P 2p, and S 2p levels for the various $MPS_3$ compounds before (dark green) and after lithium doping (bright green).** (a) – (d) show XPS measurements on the M 2p states for the bare $MPS_3$ surface as well as the lithium doped surface. The shoulder visible in the Co 2p spectrum between 770 and 780 eV originates in the $L_2M_{23}M_{45}$ auger transition, which lays close to the Co $2p_{3/2}$ peak when using an Al Kα light source. (e) and (f) present XPS measurements on the P 2p and S 2p peaks for bare $MnPS_3$ (dark green) and lithium doped $MnPS_3$ (bright green), respectively. A fit of the S 2p spectra displayed in (g) and (h) indicates only one oxidation state for the bare $MnPS_3$, while the necessity of two doublets to fit the lithium doped surface hints towards two different moieties of sulfur atoms (doped and undoped).

In contrast, Fe-, Co-, and $NiPS_3$ (**Figures 2b–d**) show substantial modifications upon lithium doping. New peaks emerge at lower binding energies relative to the original $2p_{3/2}$ and $2p_{1/2}$ peaks, suggesting the appearance of a new oxidation state rather than a uniform chemical shift. These changes evolve with lithium coverage (see **Figure S2**) and are accompanied by a reduction in spin-orbit splitting (see **Table S3**). For example, $CoPS_3$ shows a decrease in 2p peak splitting from 16 eV (bare surface) to 15 eV after doping, consistent with a transition from $Co^{2+}$ in a high-spin $3d^7$ (S = 3/2) configuration to a lower spin $3d^9$ (S = 1/2) state, potentially corresponding to $Co^0$. This interpretation is supported by comparison with previous studies of Co complexes by Frost et al. [37]. Similar, though less pronounced, reductions in spin-orbit splitting are also observed for NiPS$_3$ and FePS$_3$ (see **Table S3**).

Additionally, lithium doping alters the satellite structure of the M 2p spectra. For instance, NiPS$_3$ originally shows distinct satellites ~10 eV above the main peaks (labeled Sat2$_{3/2}$ and Sat2$_{1/2}$ in **Table S4, and SI section 5**), which vanish upon doping. Such satellite features are typically associated with ligand-to-metal charge transfer and multiplet effects involving final-state interactions [34]. Their disappearance likely reflects a change in final-state screening and electronic structure, further supporting a modified electronic configuration in the doped compounds. The remaining satellites shift in energy in accordance with the new spin-orbit splitting values.

Besides the transition metal core levels, we also analyzed the XPS signals from the phosphorus and sulfur atoms, focusing on the P 2p and S 2p regions. For Fe-, Co-, and NiPS$_3$, both peaks remain unchanged upon lithium doping, indicating that the ligand environment is largely unaffected (see **SI section 6**). The same holds for the P 2p peak in MnPS$_3$ (**Figure 2e**). However, the S 2p spectrum of MnPS$_3$ exhibits a distinct transformation upon lithium doping. As shown in **Figure 2f**, the bare surface displays the characteristic spin-orbit split 2p$_{3/2}$ and 2p$_{1/2}$ components, which are not fully resolved due to instrumental limitations. After lithium deposition, the overall intensity of the S 2p signal decreases, and an additional shoulder appears at lower binding energy. **Figures 2g and 2h** compare the S 2p spectra of the bare and Li-doped MnPS$_3$ surfaces. While the undoped sample displays a single doublet, the doped surface requires fitting with two distinct doublets. One matches the original S 2p position, while the other is shifted to lower binding energy, indicating the emergence of a second sulfur species. To summarize, the XPS data demonstrate that lithium deposition alters the oxidation state of the M$^{2+}$ ions in Fe-, Co-, and NiPS$_3$, as the additional charge is incorporated into the transition-metal d orbitals, resulting in a transition from M$^{2+}$ to a M$^0$-like configuration in CoPS$_3$. In MnPS$_3$ instead, the lithium-derived electrons predominantly localize on the P$_2$S$_6$ ligand clusters rather than the transition metal center. This contrast underscores the critical role of the transition-metal electronic configuration in determining the doping mechanism.

## ARPES study

We now investigate how the distinct electron doping behavior identified by XPS modifies the electronic band structure of the MPS$_3$ compounds. To this end, we performed angle-resolved photoemission spectroscopy (ARPES) measurements on all four systems. As a reference, we first characterized the pristine surfaces of MnPS$_3$, FePS$_3$, CoPS$_3$, and NiPS$_3$ to benchmark our DFT+U calculations and establish a basis for analyzing lithium-induced changes. The corresponding ARPES spectra along the high-symmetry path K–Γ–M are shown in **Figure 3a, c, e,** and **g**, respectively. To reduce the influence of potential effects from laser incidence angle and light polarization[38], all momentum maps were symmetrized according to the hexagonal lattice symmetry prior to extracting the band structure (see **Figure S7** for details).

The measured band structures exhibit excellent agreement with DFT calculations using the PBE+U functional, shown in **Figure 3b, d, f,** and **h**. The high photon energy of 21.2 eV used in the ARPES measurements also enables access to electronic states in neighboring Brillouin zones (BZs), presented in **Figure 3i–l**. As expected, the observable parallel momentum (k$_{||}$) range, commonly referred to as the photoemission horizon, decreases with increasing binding energy due to the reduced kinetic energy of the photoelectrons.

In all four band structures, the data is cut along the Gamma points of the central and neighboring Brillouin zones ($\Gamma_0$ and $\Gamma_1$, respectively), specifically along the path M – $\Gamma_0$ – M´ – $\Gamma_1$ – M´´ (see **Figure S7**). Notable differences in band dispersion between the two zones are observed, particularly around 3 eV and 5 eV below the valence band maximum (VBM). This effect is well known and attributed to the photoemission matrix element[38], which affects the band-specific photoemission signal strength in the first and neighboring BZs based on experimental geometry. Displaying the data from both BZs therefore facilitates a more comprehensive comparison with theoretical predictions.

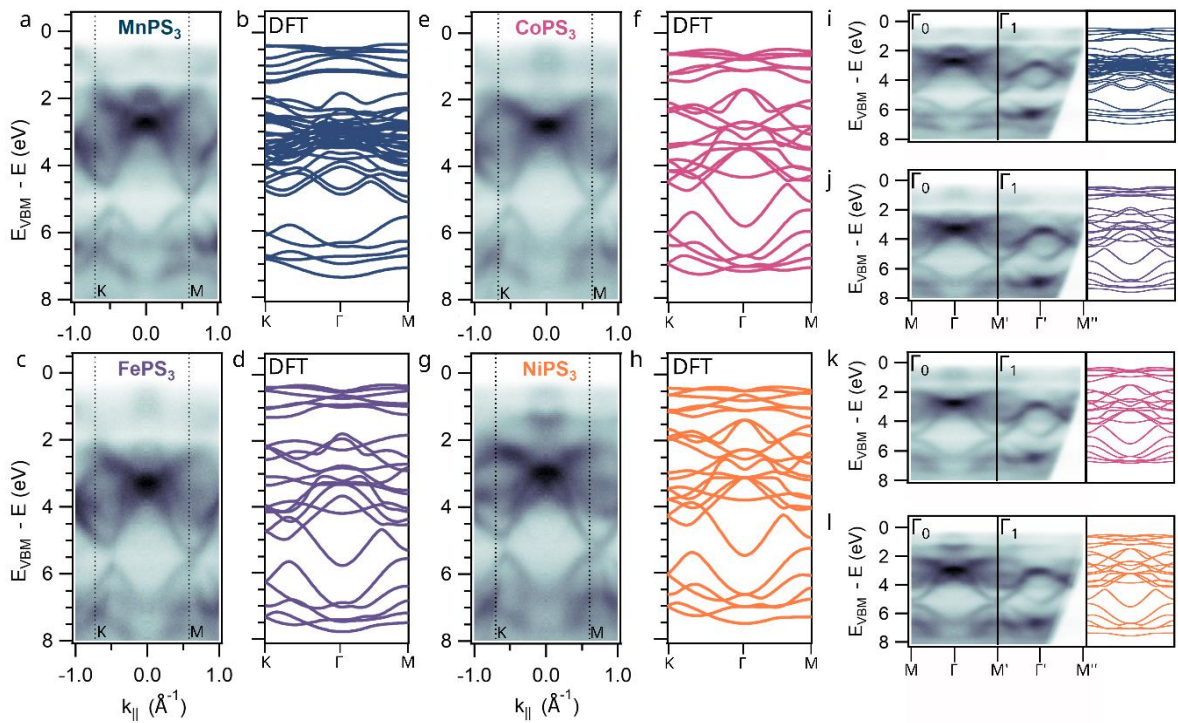

**Figure 3. ARPES measurements on different MPS$_3$ compounds and DFT+U calculations.** (a), (c) and (e), (g) show ARPES measurements of Mn-, Fe-, Co-, and NiPS$_3$ along the K – $\Gamma$ – M cut, measured with a monochromatized beam of a He II lamp at 21.2 eV photon energy. (b),(d) and (f),(h) show the corresponding DFT+U calculations along the same high symmetry path. (i) – (l) present a comparison of the ARPES intensity in the 1$^{st}$ brillouin zone ($\Gamma_0$) and the neighboring BZ ($\Gamma_1$) together with DFT derived band structures cut along M – $\Gamma$ – M' path.

In the following, we focus on the changes introduced by lithium doping in MnPS$_3$ and CoPS$_3$, which represent the two extremes in doping response within the MPS$_3$ family. **Figure 4a** and **4b** show the energy distribution curves (EDCs) for MnPS$_3$ and CoPS$_3$, respectively, before (dark green) and after (bright green) lithium deposition. In MnPS$_3$, lithium doping induces only minor changes—mainly a slight broadening of spectral features—whereas in CoPS$_3$, a clear signal emerges above the original VBM. Additionally, the peak closest to the VBM shifts to higher binding energy and appears only as a shoulder in the doped spectrum.

These experimental observations are supported by projected density of states (PDOS) calculations. For MnPS$_3$, lithium doping results in no significant redistribution of states, apart from overall spectral broadening (**Figure 4c**). The lack of energetic shifts in the Mn-related states indicates that the Mn oxidation state remains unchanged, consistent with our XPS analysis. This stability can be attributed to the half-filled $3d^5$ configuration of $Mn^{2+}$, which is a particularly energetically stable electronic configuration.

In contrast, the calculated PDOS of $CoPS_3$ (**Figure 4d**) reveals substantial changes upon lithium doping. New states appear near the VBM, which arise predominantly from hybridized Co 3d and S 3p orbitals and are shifted to lower energy in comparison to the bare surface due to Coulomb interaction with the additional electron donated by Li. Crucially, these states are not localized impurity states introduced by Li itself but rather reflect a reorganization of the $CoPS_3$ electronic structure induced by doping. This reorganization in turn implies that lithium alters the covalent Co–S interaction, potentially affecting both the electronic dispersion and magnetic exchange interactions.

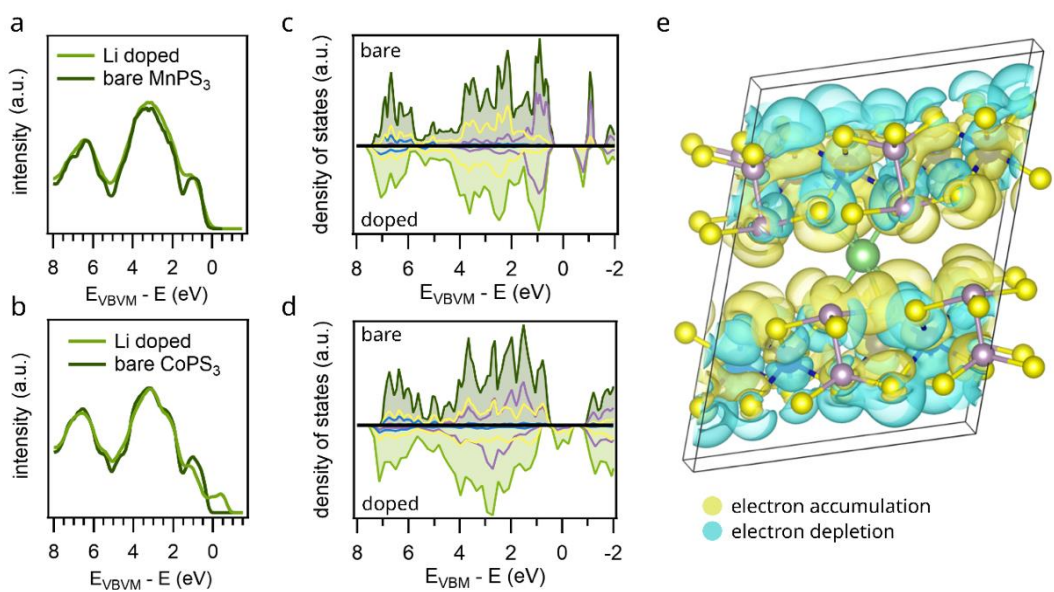

**Figure 4. Changes in the valence band and work function after alkali metal deposition.** (a), (b) display the EDC for Mn- and $CoPS_3$ before (dark green) and after alkali metal deposition (light green). (c), (d) show the PDOS calculated by DFT+U for the bare surface (top part) and doped surface (bottom part). The purple, yellow and blue lines correspond to the M, S and P ions, respectively. e) charge density difference distribution of the Li doped $CoPS_3$, where the yellow and cyan regions represent electron accumulation and depletion, respectively.

Furthermore, the Co 3d-related states near the VBM experience a substantial shift by 1.8 eV to higher binding energy. This shift signals a change in the oxidation state of Co, in agreement with the core-level shifts observed in the Co 2p XPS spectra. While this effect is also evident in the ARPES data, it is less pronounced than in theory. We attribute the discrepancy to the theoretical assumption of a full electron donated per unit cell, whereas the experimental electron doping level is likely lower. To test this interpretation, we conducted a complementary experiment using cesium instead of lithium. Due to its larger atomic radius, cesium is expected to donate fewer electrons per unit cell. As shown in **Figure S8**, the $CoPS_3$ band structure shows similar qualitative modifications upon Cs doping, but the shift of the peak near the VBM is less pronounced, confirming that the magnitude of the Co-related band shift scales with the amount of charge transferred, and reinforcing our interpretation of doping-induced oxidation state changes.

To gain deeper insight into the charge transfer mechanism upon lithium doping, we calculated the charge density difference for the Li-doped CoPS$_3$ system, shown in **Figure 4e**, using the expression:

$$\Delta\rho = \rho_{Li/CoPS_3} - \rho_{CoPS_3} - \rho_{Li}$$

where $\rho_{Li/CoPS_3}$, $\rho_{CoPS_3}$, and $\rho_{Li}$ denote the charge densities of the Li-doped system, the pristine CoPS$_3$, and an isolated Li atom, respectively. In the resulting plot, yellow and cyan regions represent electron accumulation and depletion, respectively. The data clearly show that the Li atom donates its charge to the CoPS$_3$ lattice, with the resulting electron density concentrating primarily around the Co and S atoms. This confirms charge transfer and the interaction between Li and the host lattice. To quantify the amount of transferred charge, we performed a Bader charge analysis. The results show that Li donates approximately 0.89 electrons to the CoPS$_3$ lattice. Most of this charge is delocalized over the neighboring Co and S atoms, indicating a strong interaction and redistribution of electron density within the lattice.

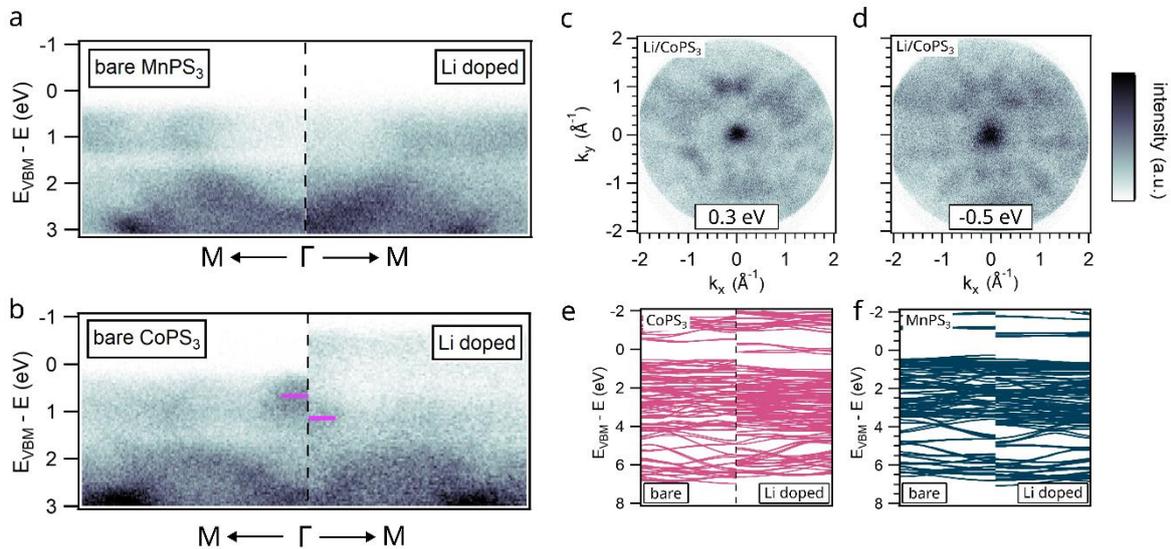

**Figure 5. Changes due to alkali metal doping investigated with ARPES.** (a) and (b) depict band structure cuts along the M - Γ – M high symmetry direction of the hexagonal BZ for Mn- and CoPS$_3$, respectively. In contrast to Figure 3, this data is not symmetrized. The left sides show the signal stemming from the clean surfaces, and the right side shows the signal from the lithium doped surfaces. The purple lines in (b) track the shift of the feature around Γ. (c) and (d) display momentum maps for Li/CoPS$_3$ at 0.3 eV below and 0.5 eV above the pristine VBM, respectively. (e) and (f) show DFT+U calculations for the clean and Li-doped CoPS$_3$ and MnPS$_3$, respectively. The calculations show the same induced behavior as extracted from the ARPES measurements with a shift of the bands close to the VBM and new states appearing in the previous band gap for CoPS$_3$, while MnPS$_3$ shows only negligible differences, except a slight shift of the VBM.

After identifying the alkali metal-induced changes using angle-integrated photoemission and PDOS calculations, momentum microscopy (MM) provides further insights into the electronic structure across the entire BZ. We begin by examining the band structure along specific high-symmetry directions. **Figure 5a and b** present ARPES cuts along the M–Γ–M path for the bare and Li-covered surfaces of MnPS$_3$ and CoPS$_3$, respectively. As discussed previously, MnPS$_3$ (**Figure 5a**) exhibits negligible changes upon lithium doping, aside from a general broadening of features and a partial closing of the gap around 1.4 eV. The overall band dispersion remains essentially unchanged within the examined energy window. In contrast, lithium doping induces pronounced modifications in the band structure of CoPS$_3$ (**Figure 5b**), the compound with the least stable d-shell configuration, particularly affecting bands derived from Co 3d orbitals. The spectral features closest to the pristine VBM, which we use as a reference energy for all spectra, shift to higher energy by approximately 400 meV, as highlighted by a purple stripe at Γ. In addition, strong spectral weight emerges up to 1 eV above the pristine VBM, with clearly dispersive features visible at both the Γ and M points of the surface Brillouin zone. This emergent occupation indicates a transition toward metallic behavior. Complementary cesium doping experiments, confirming that the extent of charge transfer into the d-shell, and its impact on the band structure, scales with the doping level (see **Figure S8**).

To understand the nature of the emerging occupation above the pristine VBM, **Figure 5c and 5d** display momentum maps (intensity vs. $k_x$, $k_y$) of Li-doped $CoPS_3$ at two selected energies: 0.3 eV below and 0.5 eV above the pristine VBM. The momentum-dependent intensity distributions demonstrates clear differences and thus confirm the dispersive character of the bands above the pristine VBM, indicating that lithium doping leads to a significant modification of the $CoPS_3$ band structure. Based on our previous analysis, we attribute this effect to be primarily driven by changes in the electronic configuration of the Co ion.

**Figure 5e** shows the DFT+U derived band structures of pristine and lithium doped $CoPS_3$. In contrast to the band structure shown in **Figure 3**, these calculations are based on the full unit cell, resulting in a greater number of visible bands. While pristine $CoPS_3$ exhibits semiconducting behavior with a finite band gap, the introduction of lithium leads to an upward shift of the Fermi level, indicative of n-type doping. This shift closes the band gap and introduces bands crossing the Fermi level, suggesting a transition from semiconducting to metallic behavior. The dispersive nature of these bands near the Fermi level agrees well with experimental observations and suggests enhanced electronic conductivity upon doping. In contrast, **Figure 5f** illustrates the calculated band structure of $MnPS_3$ with and without lithium. No significant modifications are observed, apart from a slight rigid shift to higher energy, consistent with both the ARPES and PDOS results. This behavior is the consequence of the localization of electrons on ligand orbitals, due to the high stability of its half-filled $3d^5$ shell. This suggests that charge transfer alone is insufficient to induce substantial electronic reconstruction in systems with electronically rigid d-shells.

Similar lithium and cesium doping experiments were carried out on $NiPS_3$ and $FePS_3$. Although XPS measurements indicate oxidation state changes comparable to those seen in $CoPS_3$, the ARPES data reveal markedly different electronic responses. In $FePS_3$, only very weakly dispersive states appear above the pristine VBM, while in $NiPS_3$, a pronounced increase in spectral weight is observed without clear band dispersion. These results suggest that, although charge transfer occurs, it does not induce the formation of dispersing states near the Fermi level as in $CoPS_3$. The corresponding EDCs and ARPES spectra for $FePS_3$ and $NiPS_3$ are provided in **Figure S9**.

Consistently, DFT+U calculations for Li-doped $FePS_3$ and $NiPS_3$ (**Figure S10**) reveal only minor modifications of the electronic structure, in agreement with previous theoretical work on potassium-doped $NiPS_3$ [27]. Together, these findings show that although all $MPS_3$ compounds accept electron doping to some extent, only $CoPS_3$ undergoes a significant reorganization of its low-energy band structure, highlighting its distinct chemical and electronic tunability.

# Conclusion

In this work, we combined XPS, ARPES, and DFT+U calculations to investigate how alkali metal doping modulates the electronic structure of the MPS$_3$ family of layered antiferromagnets. Our results reveal distinct and compound-specific doping mechanisms: while Fe-, Co-, and NiPS$_3$ readily accommodate additional electrons in their transition metal d-orbitals, MnPS$_3$ resists such doping due to the stability of its half-filled 3d$^5$ configuration. Instead, the donated electrons localize in the ligand environment, specifically within the P$_2$S$_6$ clusters, without significantly altering the electronic band structure near the pristine VBM. This underscores the critical role of the transition metal's electronic configuration in governing charge redistribution.

For Fe-, Co-, and NiPS$_3$, XPS reveals clear oxidation state changes in the M$^{2+}$ ions, indicating d-orbital occupation upon doping. In CoPS$_3$, ARPES measurements show a pronounced restructuring of the valence band, including a shift of Co-derived bands and the emergence of new dispersive metallic states above the pristine VBM. These modifications—supported by DFT+U calculations and corroborated by complementary Cs doping experiments—demonstrate a doping-controlled transition from semiconducting to metallic behavior and point toward a potential influence on magnetic exchange interactions.

In contrast, Li doping induces minimal changes in the experimental and theoretical band structures of MnPS$_3$, FePS$_3$, and NiPS$_3$, suggesting that charge transfer alone is insufficient to induce significant electronic reorganization in these systems. Future XMCD measurements will be crucial to elucidate the role of orbital occupation in tuning magnetic order, while time-resolved ARPES may shed light on the influence of doping on carrier dynamics—particularly in promising candidates like FePS$_3$ [12].

Overall, our study establishes alkali metal doping as a powerful, controllable method to modulate the electronic structure of 2D antiferromagnets. The ability to tune band structure and charge distribution in compounds with a less energetically stable d-shell opens new opportunities for engineering the spin and electronic properties of ionically bonded 2D materials for future spintronic and quantum technologies [2,39,40].

## Acknowledgments

The momentum microscope was financed by the DFG through the project INST 212/409 and by the "Ministerium für Kultur und Wissenschaft des Landes Nordrhein-Westfalen". We also acknowledge financial support by the DFG through project dd2D (CI 157/11-1) as well as support from the European Union's Horizon 2020 Research and Innovation Program under Project SINFONIA, grant 964396. S.B. acknowledges funding from the Volkswagen Stiftung (Momentum) through the project "dandelion" and from the Collaborative Research Center (CRC/SFB) 1375 ``NOA – Nonlinear Optics down to Atomic scales" of the German Research Foundation (DFG).

## Author contributions

*Conceptualization,* J.E.N.; *Methodology*, J.E.N., and P.B. ; *Experimental investigation*, J.E.N., T.W., P.M., and M.G.; *Experimental support*, D.J., L.S., K.S., M.S.A., V.M., and M.Ca.; *Data analysis*. J.E.N. and T.W.; *Theoretical calculations*, P.B. and S.B.; *Writing – original draft*, J.E.N., P.B., and M.C.; *Writing – review and editing*, J.E.N., P.B., T.W., P.M., D.J., L.S., K.S., V.M., M.S.A., S.B., and M.C.; *Funding acquisition*, S.B. and M.C.

## Materials availability

The raw data that support the findings of this study are available in the Zenodo database. Any additional information required to reanalyze the data reported in this paper is available from the lead contact upon request.

## Competing interests

The authors declare no competing interests.

## Declaration of generative AI and AI-assisted technologies

During the preparation of this work, the authors used ChatGPT to enhance the quality of writing by improving grammar, style, and clarity. After using this tool/service, the authors reviewed and edited the content as needed and take full responsibility for the content of the publication.

## Methods

**Sample preparation**

The MPS$_3$ crystals were grown via chemical vapor transport and purchased commercially from HQ Graphene. The exfoliation of the crystals was carried out using standard scotch-tape exfoliation in a separate chamber of an ultra-high vacuum (UHV) system at a base pressure of below 1x10$^{-9}$ mbar. Alkali metal deposition was performed using a wire-shaped dispenser from SAES Industrial, which contains the alkali metal in form of a stable salt combined with a getter material. The deposition was performed at pressures around 1x10$^{-8}$ mbar by heating the dispenser with a DC electric current of 8.5 A. Survey spectra of the various samples before and after lithium deposition are reported in **SI sections 11 and 12**.

**Momentum Microscopy and X-Ray Photoelectron Spectroscopy**

XPS measurements were conducted using a SPECS Phoibos 150 hemispherical analyzer from SPECS GmbH. The X-ray source employed in the experiments was a monochromatized Al K$\alpha$ line with a photon energy of 1486.8 eV. The beam spot size on the sample was around 2 mm in diameter. All measurements were carried out using identical source and analyzer settings: a voltage of 12.1 kV and a constant power of 111 W. As the samples investigated are semiconductors, the Fermi edge used to calibrate the binding energy scale was extracted from measurements on an Au(111) single crystal. XPS spectra were analyzed with the XPS Tools (XPST)[41] package for Igor Pro. Peak fitting was based on a Gauss-Lorentzian sum function to approximate a Voigt profile and incorporates a Shirley background. The Gauss-Lorentzian ratio was set to 0.3.

The measurements of the valence band structure were taken with the Kreios 150 MM momentum microscope from SPECS GmbH[42,43], coupled to a UVS 300 UV light source with a monochromator and a fs-XUV source[44]. For data acquisition, the He 1$\alpha$ line with a photon energy of 21.22 eV and the 10$^{th}$ harmonic of the XUV source at 26.4 eV were used, both with p-polarization. All data were acquired at room temperature (300 K), corresponding to the paramagnetic phase of the different MPS$_3$ compounds. The instrument allows different reciprocal space magnifications, the lowest of which results in an accessible photoelectron parallel momentum of up to ± 3 Å$^{-1}$. For the reported momentum maps, the second lowest k-magnification was used, enabling a total available parallel momentum of ±2.2 Å$^{-1}$.

**Density functional theory (DFT)**

All DFT[45,46] calculations were performed using the Vienna Ab initio Simulation Package (VASP)[47,48]. Ion-electron interactions for all elemental constituents were described using the projector augmented wave (PAW) method as implemented in VASP. The Perdew-Burke-Ernzerhof (PBE) exchange-correlation functional[49] under the generalized gradient approximation (GGA) was employed. Structural optimization was carried out using the PBE+U functional, incorporating the Hubbard U parameters[50] to accurately describe the strongly correlated systems. Empirical U values of 1.9[11], 4.5, 4.6, and 5.0 eV[51] were used for Fe, Ni, Co, and Mn, respectively. The U values were derived by comparison of the experimentally obtained band structures to calculations for different U values (see **Figure S13**). Self-consistent calculations were performed with PBE+U along with van der Waals (vdW) corrections, relaxing all ions until the Hellmann-Feynman forces were less than 0.001 eV/Å. The two-body vdW interaction, as devised by Tkatchenko-Scheffler, was included during optimization. A kinetic energy cut-off of 600 eV was used for the plane-wave basis set, ensuring total energy convergence within 10-5 eV. A Γ-centered 6×6×6 k-point grid was used to sample the Brillouin zone, unless stated otherwise. A vacuum spacing of 15 Å was added to eliminate spurious interactions between periodic images. The nonmagnetic (NM) case was assumed by neglecting spin degrees of freedom.

**Formation energy**

The crystal structures of all transition metal phosphorus trichalcogenides ($MPS_3$; M = Fe, Ni, Co, Mn) were taken from experimental data in the paramagnetic phase. These compounds crystallize in a monoclinic layered structure with the C2/m space group (No. 12). The Li atom was doped into the 2×2×2 supercells at different sites and optimized using the PBE+U functional. The most stable geometry was considered where the Li atom is intercalated between the layers. Subsequently, to assess the stability of these compounds, the formation energies[52] were calculated using the following equation:

$$E_f = E_{defect} - E_{pristine} - n_{Li}\mu_{Li}$$

where $E_f$, $E_{defect}$, $E_{pristine}$, $n_{Li}$ and $\mu_{Li}$ represent the defect formation energy, total DFT energy of the doped system, DFT energy of the pristine system, number of Li atoms, and the chemical potential of Li, respectively. The computed formation energies of $FePS_3$, $NiPS_3$, $CoPS_3$ and $MnPS_3$ are -1.39, -2.07, -3.71, -1.49 eV, respectively, suggesting that all these compounds are thermodynamically stable.

# References

[1] K. S. Burch, D. Mandrus, J.-G. Park, *Nature* **2018**, *563*, 47.

[2] H. Zhong, D. Z. Plummer, P. Lu, Y. Li, P. A. Leger, Y. Wu, *Mater. Quantum Technol.* **2025**, *5*, 012001.

[3] L. Sternemann, D. M. Janas, J. E. Nitschke, K. Schiller, T. Willershausen, L. Becker, A. Isaeva, G. Zamborlini, S. Tappertzhofen, M. Cinchetti, *2D Mater.* **2025**, DOI 10.1088/2053-1583/add7ea.

[4] J. Yang, Y. Zhou, Q. Guo, Y. Dedkov, E. Voloshina, *RSC Adv.* **2020**, *10*, 851.

[5] P. Wang, F. Lian, R. Du, X. Cai, S. Bao, Y. Han, J. Xiao, K. Watanabe, T. Taniguchi, J. Wen, H. Yang, A. S. Mayorov, L. Wang, G. Yu, *Appl. Phys. Lett.* **2024**, *124*, 012406.

[6] B. Huang, G. Clark, D. R. Klein, D. MacNeill, E. Navarro-Moratalla, K. L. Seyler, N. Wilson, M. A. McGuire, D. H. Cobden, D. Xiao, W. Yao, P. Jarillo-Herrero, X. Xu, *Nat. Nanotechnol.* **2018**, *13*, 544.

[7] S. Selter, Y. Shemerliuk, M.-I. Sturza, A. U. B. Wolter, B. Büchner, S. Aswartham, *Phys. Rev. Mater.* **2021**, *5*, 073401.

[8] J. M. Pereira, D. Tezze, M. Ormaza, L. E. Hueso, M. Gobbi, *Adv. Phys. Res.* **2023**, *2*, DOI 10.1002/apxr.202200084.

[9] Y. Liu, Z. Wang, G. Hu, X. Chen, K. Xu, Y. Guo, Y. Xie, C. Wu, *Precis. Chem.* **2025**, *3*, 51.

[10] F. Mertens, D. Mönkebüscher, U. Parlak, C. Boix-Constant, S. Mañas-Valero, M. Matzer, R. Adhikari, A. Bonanni, E. Coronado, A. M. Kalashnikova, D. Bossini, M. Cinchetti, *Adv. Mater.* **2023**, *35*, e2208355.

[11] J. E. Nitschke, D. L. Esteras, M. Gutnikov, K. Schiller, S. Mañas-Valero, E. Coronado, M. Stupar, G. Zamborlini, S. Ponzoni, J. J. Baldoví, M. Cinchetti, *Mater. Today Electron.* **2023**, *6*, 100061.

[12] J. E. Nitschke, L. Sternemann, M. Gutnikov, K. Schiller, E. Coronado, A. Omar, G. Zamborlini, C. Saraceno, M. Stupar, A. M. Ruiz, D. L. Esteras, J. J. Baldoví, F. Anders, M. Cinchetti, *Newton* **2025**, 100019.

[13] J. Strasdas, B. Pestka, M. Rybak, A. K. Budniak, N. Leuth, H. Boban, V. Feyer, I. Cojocariu, D. Baranowski, J. Avila, P. Dudin, A. Bostwick, C. Jozwiak, E. Rotenberg, C. Autieri, Y. Amouyal, L. Plucinski, E. Lifshitz, M. Birowska, M. Morgenstern, *Nano Lett.* **2023**, *23*, 10342.

[14] B. Pestka, J. Strasdas, G. Bihlmayer, A. K. Budniak, M. Liebmann, N. Leuth, H. Boban, V. Feyer, I. Cojocariu, D. Baranowski, S. Mearini, Y. Amouyal, L. Waldecker, B. Beschoten, C.

Stampfer, L. Plucinski, E. Lifshitz, P. Kratzer, M. Morgenstern, *ACS Nano* **2024**, DOI 10.1021/acsnano.4c12520.

[15] Y. Dedkov, Y. Guo, E. Voloshina, *Electron. Struct.* **2023**, *5*, 043001.

[16] C. Autieri, G. Cuono, C. Noce, M. Rybak, K. M. Kotur, C. E. Agrapidis, K. Wohlfeld, M. Birowska, *J. Phys. Chem. C* **2022**, *126*, 6791.

[17] A. Harchol, S. Zuri, E. Ritov, F. Horani, M. Rybak, T. Woźniak, A. Eyal, Y. Amouyal, M. Birowska, E. Lifshitz, *2D Mater.* **2024**, *11*, 035010.

[18] N. Chakraborty, A. Harchol, A. Abu-Hariri, R. K. Yadav, M. Dawod, D. P. Gaonkar, K. Sharma, A. Eyal, Y. Amouyal, D. Naveh, E. Lifshitz, *arXiv* **2024**, DOI 10.48550/arxiv.2410.11264.

[19] R. Basnet, A. Wegner, K. Pandey, S. Storment, J. Hu, *Phys. Rev. Mater.* **2021**, *5*, 064413.

[20] R. Basnet, T. Patel, J. Wang, D. Upreti, S. K. Chhetri, G. Acharya, M. R. U. Nabi, J. Sakon, J. Hu, *Adv. Electron. Mater.* **2024**, DOI 10.1002/aelm.202300738.

[21] R. Basnet, D. Ford, K. TenBarge, J. Lochala, J. Hu, *J. Phys.: Condens. Matter* **2022**, *34*, 434002.

[22] R. Basnet, A. Wegner, K. Pandey, S. Storment, J. Hu, *Phys. Rev. Mater.* **2021**, *5*, 064413.

[23] D. Upreti, R. Basnet, M. M. Sharma, S. K. Chhetri, G. Acharya, M. R. U. Nabi, J. Sakon, B. Da, M. Mortazavi, J. Hu, *J. Phys.: Condens. Matter* **2025**, *37*, 135805.

[24] X. Li, X. Wu, J. Yang, *J. Am. Chem. Soc.* **2014**, *136*, 11065.

[25] R. Clement, L. Lomas, J. P. Audiere, *Chem. Mater.* **1990**, *2*, 641.

[26] D. Chen, C. Wang, C. Peng, *Phys. Chem. Chem. Phys.* **2024**, *26*, 8436.

[27] Y. Cao, Q. Tan, Y. Guo, C. G. Vieira, M. S. C. Mazzon, J. Laverock, N. Russo, H. Gao, C. Jozwiak, A. Bostwick, E. Rotenberg, J. Guo, M. Yi, M. J. S. Matos, X. Ling, K. E. Smith, *arXiv* **2024**.

[28] X. Wu, Z. Shen, W. Xiao, J. Yang, C. Song, *J. Mater. Sci.: Mater. Electron.* **2022**, *33*, 1871.

[29] R. P. Gupta, S. K. Sen, *Phys. Rev. B* **1974**, *10*, 71.

[30] R. P. Gupta, S. K. Sen, *Phys. Rev. B* **1975**, *12*, 15.

[31] I. Brotons-Alcázar, R. Torres-Cavanillas, M. Morant-Giner, M. Cvik, S. Mañas-Valero, A. Forment-Aliaga, E. Coronado, *ChemRxiv* **2021**, DOI 10.26434/chemrxiv-2021-lph4p.

[32] A. P. Grosvenor, B. A. Kobe, M. C. Biesinger, N. S. McIntyre, *Surf. Interface Anal.* **2004**, *36*, 1564.

[33] E. S. Ilton, J. E. Post, P. J. Heaney, F. T. Ling, S. N. Kerisit, *Appl. Surf. Sci.* **2016**, *366*, 475.

[34] Q. Lu, *ACS Nano* **2024**, *18*, 13973.

[35] Yu. G. Borod'ko, S. I. Vetchinkin, S. L. Zimont, I. N. Ivleva, Yu. M. Shul'ga, *Chem. Phys. Lett.* **1976**, *42*, 264.

[36] F. A. Stevie, C. L. Donley, *J. Vac. Sci. Technol. A: Vac., Surf., Films* **2020**, *38*, 063204.

[37] D. C. Frost, C. A. McDowell, I. S. Woolsey, *Chem. Phys. Lett.* **1972**, *17*, 320.

[38] S. Moser, *J. Electron Spectrosc. Relat. Phenom.* **2017**, *214*, 29.

[39] M. C. Lemme, D. Akinwande, C. Huyghebaert, C. Stampfer, *Nat. Commun.* **2022**, *13*, 1392.

[40] X. Jiang, Q. Liu, J. Xing, N. Liu, Y. Guo, Z. Liu, J. Zhao, *Appl. Phys. Rev.* **2021**, *8*, 031305.

[41] M. Schmid, H. Steinrück, J. M. Gottfried, *Surf. Interface Anal.* **2014**, *46*, 505.

[42] S. Ponzoni, F. Paßlack, M. Stupar, D. M. Janas, G. Zamborlini, M. Cinchetti, *Adv. Phys. Res.* **2023**, *2*, DOI 10.1002/apxr.202200016.

[43] D. M. Janas, A. Windischbacher, M. S. Arndt, M. Gutnikov, L. Sternemann, D. Gutnikov, T. Willershausen, J. E. Nitschke, K. Schiller, D. Baranowski, V. Feyer, I. Cojocariu, K. Dave, P. Puschnig, M. Stupar, S. Ponzoni, M. Cinchetti, G. Zamborlini, *Inorg. Chim. Acta* **2023**, *557*, 121705.

[44] K. J. Schiller, L. Sternemann, M. Stupar, A. Omar, M. Hoffmann, J. E. Nitschke, V. Mischke, D. M. Janas, S. Ponzoni, G. Zamborlini, C. J. Saraceno, M. Cinchetti, *Sci. Rep.* **2025**, *15*, 3611.

[45] P. Hohenberg, W. Kohn, *Phys. Rev.* **1964**, *136*, B864.

[46] W. Kohn, L. J. Sham, *Phys. Rev.* **1965**, *140*, A1133.

[47] G. Kresse, J. Furthmüller, *Comput. Mater. Sci.* **1996**, *6*, 15.

[48] G. Kresse, D. Joubert, *Phys. Rev. B* **1998**, *59*, 1758.

[49] J. P. Perdew, K. Burke, M. Ernzerhof, *Phys. Rev. Lett.* **1996**, *77*, 3865.

[50] E. H. Lieb, *Phys. Rev. Lett.* **1989**, *62*, 1201.

[51] Q. Pei, Y. Song, X. Wang, J. Zou, W. Mi, *Sci. Rep.* **2017**, *7*, 9504.

[52] P. Bhumla, M. Kumar, S. Bhattacharya, *Nanoscale Adv.* **2020**, *3*, 575.